\documentclass[a4paper,11pt]{article}
\usepackage{pos}
\usepackage{siunitx}
\usepackage{physics}
\usepackage{hyperref}
\usepackage{placeins}
\newcommand{\IM}{\mathrm{Im}}
\title{Search for a Lee-Yang edge singularity in high-statistics Wuppertal-Budapest data}

\author*[a]{Alexander Adam}
\author[a]{Szabolcs Bors\'anyi}
\author[a,b,c,d]{Zoltán Fodor}

\author[a]{Jana N. Guenther}

\author[e]{Paolo Parotto}
\author[d]{Attila P\'asztor}
\author[d,f]{D\'avid Peszny\'ak}
\author[a]{Ludovica Pirelli}
\author[a]{Chik Him Wong}

\affiliation[a]{University of Wuppertal, Department of Physics, Wuppertal D-42119, Germany}

\affiliation[b]{Pennsylvania State University, Department of Physics, State College, PA 16801, USA}

\affiliation[c]{Jülich Supercomputing Centre, Forschungszentrum Jülich, Jülich D-542425, Germany}
\affiliation[d]{E{\"o}tv{\"o}s University, Budapest 1117, Hungary}

\affiliation[e]{Dipartimento di Fisica, Università di Torino and INFN Torino, Via P. Giuria 1, I-10125 Torino, Italy}
\affiliation[f]{HUN-REN Wigner Research Centre for Physics, Budapest 1121, Hungary}

\emailAdd{alexander.adam@uni-wuppertal.de}

\abstract{Near a critical endpoint the Lee-Yang edge singularity approaches the real axis in the complex chemical potential plane. In the vicinity of the critical point the functional form of this approach depends on the universality class. Assuming a three dimensional Ising critical point in the QCD phase diagram the location of the critical endpoint can be extrapolated provided that the position of the Lee-Yang edge singularity is known at multiple temperatures. A popular method to estimate the position of a singularity is to model the free energy as a rational function of the baryon chemical potential $\mu_\text{B}$. The parameters of this model can be constrained by the cumulants of the net baryon density taken at $\mu_\text{B}^2\leq0$
. Using high-statistics simulations on a lattice $16^3\times8$ by the Wuppertal-Budapest Collaboration we estimate the location of the closest singularity in the QCD phase diagram. We also compare various models for the functional form of the free energy and discuss the predictive power of this approach.}

\FullConference{The 41st International Symposium on Lattice Field Theory (LATTICE2024)\\
 28 July - 3 August 2024\\
Liverpool, UK\\}

\begin{document}
\maketitle

\section{Introduction}
The QCD phase diagram at finite temperature and baryon chemical potential ($\mu_\text{B}$) is a central focus of modern nuclear and particle physics, with profound implications for understanding heavy-ion collision experiments and the properties of compact stars. 
Using lattice simulations to investigate QCD allows for first-principles solutions with systematically improvable errors.
At $\mu_\text{B} = 0$, lattice QCD has firmly established that the transition from the confined hadronic phase to the deconfined quark-gluon plasma (QGP) phase is a smooth crossover \cite{Aoki:2006we}. 
However, at finite $\mu_\text{B}$, the nature of the transition is expected to change, potentially culminating in a critical endpoint (CEP) where the crossover transitions into a first-order phase transition \cite{Du:2024wjm}.

The search for this CEP and to understand the thermodynamics of strongly interacting matter under extreme conditions is a great challenge for lattice QCD.
Direct simulations at finite $\mu_\text{B}$ are hindered by the sign problem. To circumvent this, various approaches have been developed.
A popular method is the analytical continuation of thermodynamical observables from imaginary to real values of the chemical potential.
Similar techniques relies on higher order derivatives taken at $\mu_\text{B} = 0$.
All these approaches carry a model dependence due to the ambiguity in the choice of the ansatz. 

A popular way to search for criticality is the determination of the Lee-Yang zeros (LYZs), the complex valued chemical potential where the partition sum vanishes \cite{Yang:1952be}.
While a Lee-Yang zero (LYZ) will hit the real axis only in the case of a true transition, its approach to zero imaginary value is governed by universal scaling \cite{Skokov:2024fac}.

This was exploited in Refs.~\cite{Bollweg:2022rps,Basar:2023nkp,Clarke:2024ugt} using state of the art lattice simulations to constrain the position of the LYZ.
Beyond lattice QCD LYZ were explored in a diagrammatic context in Ref.~\cite{Wan:2024xeu}.

In this work we explore this class of methods for the search of the CEP using a high precision data set of the Wuppertal-Budapest collaboration. 
\section{Method}
\begin{figure}[ht]
    \centering
    \includegraphics[width=0.85\linewidth]{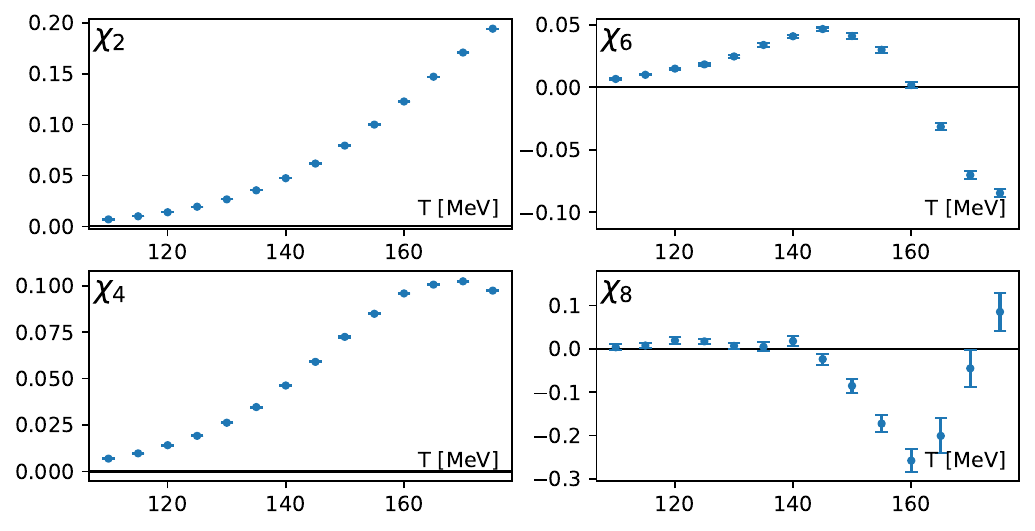}
    \caption{The baryon susceptibilities used in the analysis of the Lee-Yang edge singularity singularity plotted against the temperature range of \SIrange{110}{175}{\MeV}}
    \label{fig:chi_plot}
\end{figure}

To investigate the zeros of the partition function $ Z $, commonly referred to as Lee-Yang zeros , we analyze the logarithm of the partition function, which corresponds to the pressure $ \log(Z) = p/T^4 $. The zeros of the partition function manifest as singularities in the pressure, as the logarithmic function diverges at $ Z = 0 $.
Expanding the pressure around $ \mu_B = 0$ for a fixed temperature $ T $ results in the following Taylor expansion for $ \Delta \hat p $
\begin{equation}
\Delta \hat p = \frac{p(T, \mu_B) - p(T, 0)}{T^4} = \sum_{n=1}^\infty \frac{\chi_{2n}(T)}{(2n)!} \qty(\frac{\mu_B}{T})^{2n},
\label{eq:deltaptaylor}
\end{equation}
where the coefficients $ \chi_{2n}(T) $ are defined as:
\begin{equation}
\chi_n(T) = \frac{\partial^n (p / T^4)}{\partial (\mu_B / T)^n}.
\end{equation}
These coefficients, $ \chi_{2n}(T) $, are determined through Monte Carlo simulations at the specified temperature. 
The simulations are carried out on a $16^3 \times 8$ lattice with physical quark masses, employing 4 steps of HEX smearing. 
For each temperature, the number of configurations is on the order of $\mathcal{O}(2 \times 10^6)$. 
The simulated values of $ \chi_{2n}(T) $ are shown in \autoref{fig:chi_plot} over the temperature range included in the calculation of the LYZs.
Using these simulated coefficients of the Taylor expansion (\autoref{eq:deltaptaylor}) can provide insight into the proximity of singularities by determining the radius of convergence (Refs. \cite{Giordano:2019slo,Mukherjee:2019eou}).
In this work we will search for the Lee-Yang edge singularity by determining the LYZs.
The LYZs themselves are estimated by using a rational approximation of observables and calculating their point of divergence by identifying the roots of the denominator.
To match the Taylor and the rational approximation one can algebraically transform the coefficients of the Taylor series into the coefficients of a rational polynomial approximation, known as a Padé approximation, of corresponding order.
The following \autoref{eq:taylor_and_pade} depicts a general Taylor and Padé approximation.
\begin{equation}
    T_l(x)=\sum_{i=0}^l c_i x^i \qquad \text{Padé}[m,n]:\frac{P_m(x)}{1+Q_n(x)}=\frac{\sum_{i=0}^m a_i x^i}{1+\sum_{i=1}^n b_i x^i}
    \label{eq:taylor_and_pade}
\end{equation}
The algebraic transformation can be constructed by equating the Taylor series $T_l(x)$ to the Padé approximation of the form $P_m(x)/(1+Q_n(x))$
which can be rearranged as  
\begin{equation}
    P_m(x) = T_l(x) (1 + Q_n(x)).
\end{equation}  
The coefficients of the Padé approximation are then determined by matching the appropriate number of derivatives at the expansion point of the Taylor series.  
The Padé approximations used in the later analysis for pressure, density, and susceptibility are given, where $\hat\mu = \qty(\frac{\mu_B}{T})$.
\begin{align}
    \Delta p (T):\qquad \frac{a_1 \hat\mu^2 + a_2 \hat\mu^4}{1+b_1 \hat\mu^2+b_2\hat\mu^4} &\stackrel{!}{=} \sum_{i=1}  \frac{\chi_{2i}}{(2i)!} \hat\mu^{2i} \label{eq:pade_deltap}\\
    \chi_1(T):\qquad \frac{a_1 \hat\mu^1 + a_2 \hat\mu^3}{1+b_1 \hat\mu^2+b_2\hat\mu^4} &\stackrel{!}{=} \sum_{i=0} \frac{\chi_{2i+1}}{(2i+1)!}\hat\mu^{2i+1}  \label{eq:pade_chi1}\\
    \chi_2(T):\qquad \frac{a_1 \hat\mu^0 + a_2 \hat\mu^2}{1+b_1 \hat\mu^2+b_2\hat\mu^4} &\stackrel{!}{=} \sum_{i=0} \frac{\chi_{2i+2}}{(2i!)}\hat\mu^{2i}
    \label{eq:pade_chi2}
\end{align}
Mathematically, these correspond to $[4,4]$, $[3,4]$, and $[2,4]$ Padé approximations in $\hat{\mu}$, respectively. However, each has only two nonzero coefficients in both the numerator and denominator.  
The coefficients of the odd exponents in the pressure and susceptibility, as well as the even exponents in the density, are set to zero due to the symmetry of the respective functions. Additionally, the leading constant coefficient of the Padé approximation for the pressure is set to zero, as the pressure difference is being analyzed.  
Since they share the same free coefficients as a $[1,2]$ Padé, they are referred to as such in the following.  

As an example, the analytical solutions defining the algebraic system to be solved for the [1,2]-Padé using the susceptibility as the observable to be fitted then read as follows
\begin{equation}
\eval{\dv[2n]{\hat\mu}}_{\hat\mu=0} (a_1 + a_2 \hat\mu^2)= \eval{\dv[2n]{\hat\mu}}_{\hat\mu=0}(\chi_2+\frac{\chi_4}{2!} \hat\mu^2+\frac{\chi_6}{4!} \hat\mu^4 + \frac{\chi_8}{6!} \hat\mu^6)(1+b_1 \hat\mu^2+b_2\hat\mu^4)
\end{equation}
where $n$ ranges from zero to 3. 

Finally, the roots of $1 + Q_n(x)$ serve as a proxy for the location of the Lee-Yang zeros in the complex $\mu$ plane.
The location of the Lee-Yang edge singularity can be extracted from the location of the Lee-Yang zeros by extrapolating the LYZ to the point where its imaginary part vanishes. 
This extrapolation can be performed using a scaling relation.
Here the universal scaling of the 3D Ising model is assumed.
As for the scaling variable, multiple choices are possible leading to different ansatzes, which should align when sufficiently close to the Lee-Yang edge singularity. 
\autoref{eq:abc} presents a general ansatz, where $f$ denotes a function resulting in a scaling variable that depends on the baryon chemical potential of the estimated Lee-Yang zero, $\mu_B^{\text{LYZ}}$, and its corresponding temperature, $T^{\text{LYZ}}$.
\begin{equation}
\IM[f(\mu_B^{\text{LYZ}}, T)] = \kappa \left( \Delta T \right)^{\beta \delta}
\label{eq:abc}
\end{equation}  
Furthermore, $\kappa$ is a constant, $\Delta T = T - T_c$, and $\beta$ and $\delta$ are the critical exponents of the 3D Ising model.  
As later on in the analysis we want to perform linear fits with temperature on the $x$-axis, we rescale the $y$-axis.
This is equivalent to rewriting \autoref{eq:abc} into a form of $\alpha\cdot x=y$:
\begin{equation}
    \kappa \Delta T = \IM(f(\mu_B^{\text{LYZ}}, T)^{1/\beta \delta}.
\end{equation}
In this formulation, the intersection of a linear fit with the $x$-axis corresponds to the temperature of the Lee-Yang edge singularity, which is the critical temperature $T_c$.  
For scaling observables we use:
\begin{equation}
    \mu,\frac{\mu}{T},\mu^2,\frac{\mu^2}{T^2}
\end{equation}

\FloatBarrier\section{Analysis}

\begin{figure}[ht]
    \centering
    \begin{minipage}{0.5\textwidth}
        \centering
    \includegraphics[width=\textwidth]{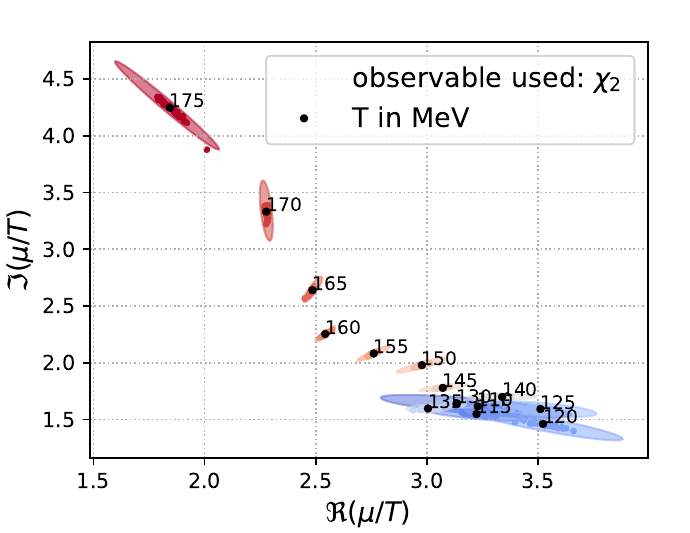}
\end{minipage}\begin{minipage}{0.5\textwidth}
    \centering
    \includegraphics[width=\textwidth]{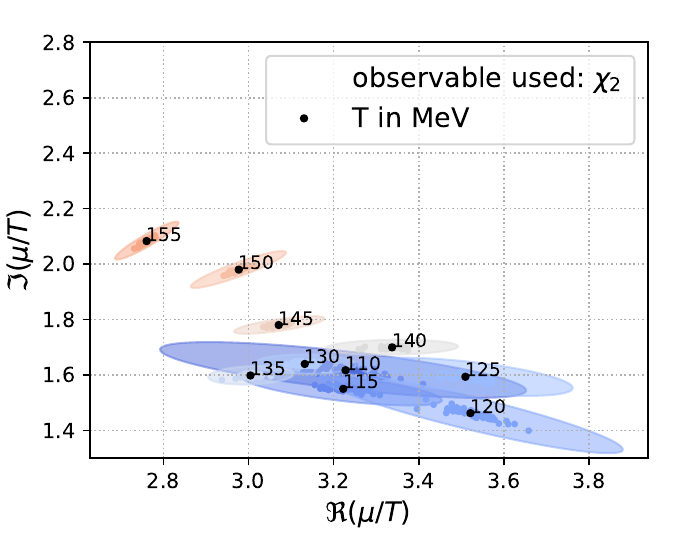}
\end{minipage}
    \caption{The location of the LYZs extracted using a Padé ansatz for the susceptibility $\chi_2(T)$, see \autoref{eq:pade_chi2}. The left plot shows the entire temperature range of \SIrange{110}{175}{\MeV} while the right plot depicts a reduced range of \SIrange{110}{155}{\MeV}. The ellipses represent the 68\% confidence region based on the covariance matrix computed from the jackknife samples.}
    \label{fig:sings-chi2}
\end{figure}

\begin{figure}
\begin{minipage}{0.5\textwidth}
        \centering
    \includegraphics[width=\textwidth]{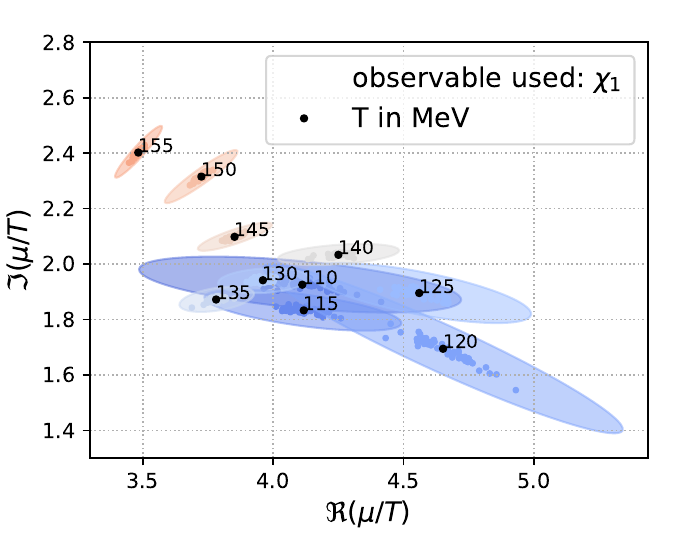}
\end{minipage}\begin{minipage}{0.5\textwidth}
    \centering
    \includegraphics[width=\textwidth]{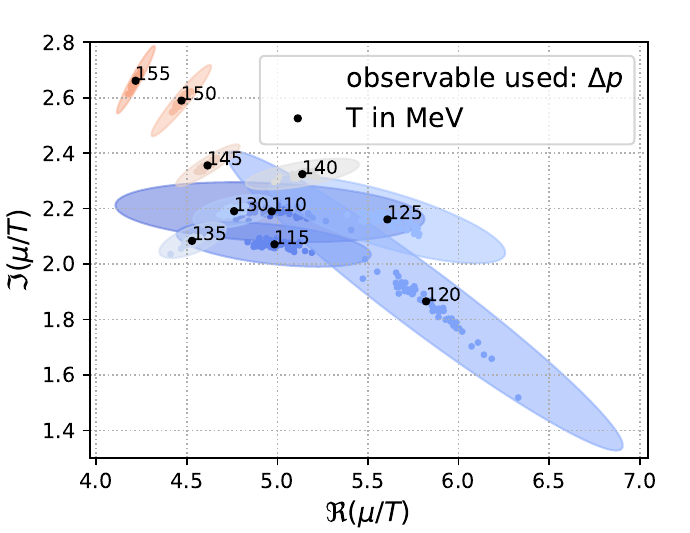}
\end{minipage}

\caption{Location of the LYZs in a reduced range of \SIrange{110}{155}{\MeV}, each extracted with a [1,2] Padé ansatz with the trivial lower order coefficients in the nominator removed. The left plot uses the density $\chi_1(T)$ as the underlying observable, while the right plot uses the pressure difference $\Delta p(T)$ as the underlying observable. See \autoref{eq:pade_deltap} and \autoref{eq:pade_chi1} for the specific Padé used.}
\label{fig:sings-deltap_chi1}
\end{figure}

In order to visualize the Lee-Yang zeros, the location of the singularities is typically plotted by visualizing the poles and zeros of the Padé approximation \cite{Singh:2022diw}. This approach allows for the identification of potential overlaps and provides a check for cancellations, such as removable singularities. 
However, due to the simplicity of the $[1,2]$-Padé approximations used here, the poles and zeros never align, thus the plotting of the zeros can be omitted. 
\autoref{fig:sings-chi2} shows the estimated locations of the Lee-Yang zeros for all temperatures based on the Padé approximation of the susceptibility $\chi_2(T)$. 
Statistical errors are estimated using jackknife resampling with 48 bins.
The plot on the left displays the locations for all temperatures in the range of \SIrange{110}{175}{\MeV}, while the plot on the right shows only the temperature range \SIrange{110}{155}{\MeV}. 
The black dot represents the central sample, and the colored points represent the estimated poles for each jackknife sample. The colored ellipses indicates the 68\% confidence region, derived from the covariance matrix obtained through the jackknife analysis.
Repeating the analysis using $\chi_1(T)$ and $\Delta \hat p(T)$ as the underlying observables with their respective Padé formulations, as specified in \autoref{eq:pade_chi1} and \autoref{eq:pade_deltap}, on the reduced range of \SIrange{110}{155}{\MeV} yields \autoref{fig:sings-deltap_chi1}.

The imaginary axis in all plots of the reduced range has been fixed for comparison, while the real axis has been allowed to vary. This choice is motivated by the fact that the imaginary part dominates the subsequent determination of $T_c$.
\newcommand\tempvarscalingABC{1}

\begin{figure}
    \begin{minipage}{0.5\textwidth}
        \centering
    \includegraphics[width=\tempvarscalingABC\textwidth]{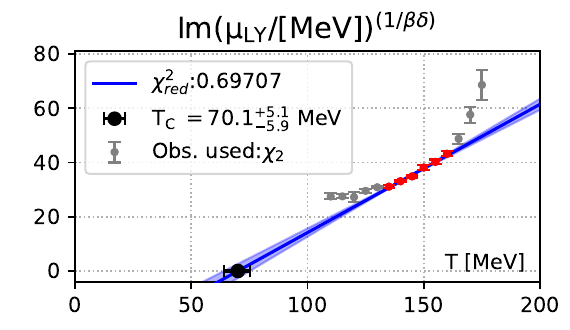}
    \end{minipage}\begin{minipage}{0.5\textwidth}
        \centering
    \includegraphics[width=\tempvarscalingABC\textwidth]{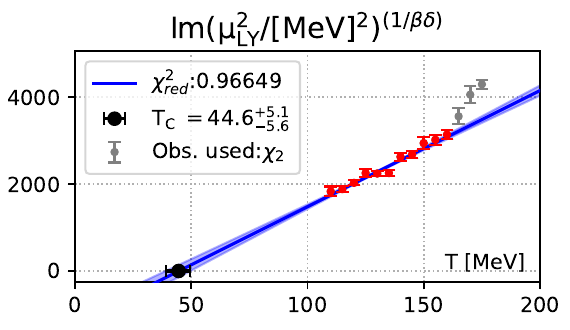}
    \end{minipage}
    \begin{minipage}{0.5\textwidth}
        \centering
    \includegraphics[width=\tempvarscalingABC\textwidth]{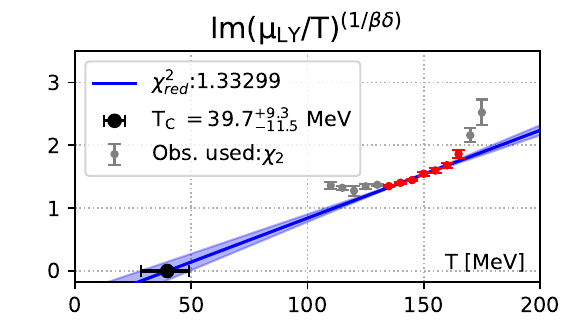}
    \end{minipage}\begin{minipage}{0.5\textwidth}
        \centering
    \includegraphics[width=\tempvarscalingABC\textwidth]{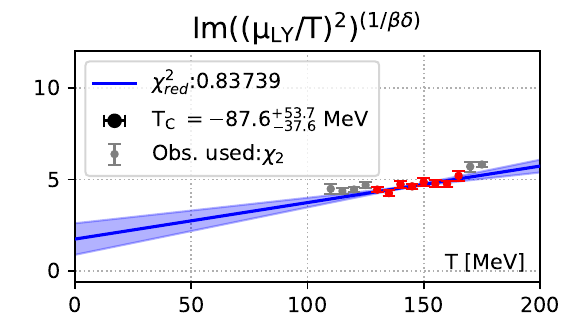}
    \end{minipage}

    \caption{Selected scaling fits for the imaginary part of the Lee-Yang zeros are shown, using different scaling ansatzes across various data ranges. Here, the susceptibility $\chi_2(T)$ is employed for the Padé ansatz. The red points indicate the LYZs used for the fit while the gray point represent the the remaining LYZs} 
    \label{fig:comp_ansatz}
\end{figure}        
Another source of systematic uncertainty arises from the choice of the scaling ansatz used to extrapolate the Lee-Yang edge singularity from the Lee-Yang zeros. 
\autoref{fig:comp_ansatz} illustrates the results, for the four different scaling variables.
For each plot the LYZs obtained are from the Padé approximation of the susceptibility.
Each ansatz is applied to an individually selected fit range, ensuring good fit quality for demonstration purposes.
The red points indicate the LYZs used for the fit, while the gray points represent the remaining LYZs.
\begin{figure}
    \centering
    \includegraphics[width=0.85\linewidth]{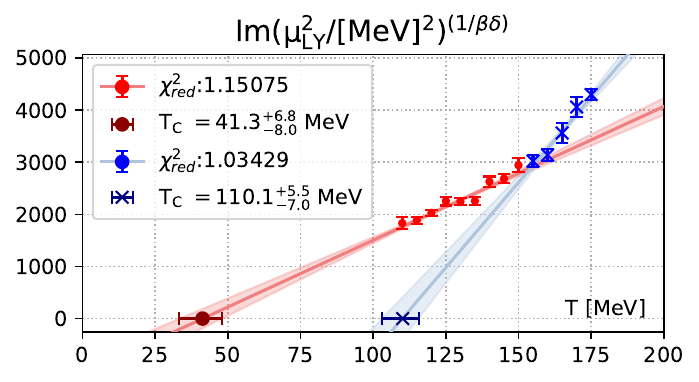}
    \caption{Scaling fits for the imaginary part of the LYZs using the scaling ansatz $\IM(\mu_{LY}^2)^{1/\beta\gamma}$ showing systematic variation with different temperature selections for the extrapolation: red (\SIrange{110}{150}{\MeV}), blue (\SIrange{155}{175}{\MeV}).}
    \label{fig:comp_range}
\end{figure}
It is evident that different approaches significantly influence both the shape and the resulting location of the Lee-Yang edge singularity.
To demonstrate the influence of the fit range, \autoref{fig:comp_range} presents two extrapolations with the scaling of $\mu_B^2$, where the only difference arises from the use of different fit ranges.  
The red points represent the LYZs obtained from temperatures in the range of \SIrange{110}{150}{\MeV}, while the blue points correspond to the remaining LYZs in the range of \SIrange{155}{175}{\MeV}.  
When considering the locations of all LYZs in the comparison of the ansatzes, as shown in \autoref{fig:comp_ansatz}, it becomes evident that higher temperatures, with their steeper slopes, will systematically lead to higher values for $T_c$ in the extrapolations.
It appears that this change in the slope occurs approximately at the crossover temperature of $\approx \SI{160}{\MeV}$.
A similar effect can be observed in a scaling plot presented in reference \cite{Wan:2024xeu}, which was obtained using Dyson-Schwinger equations.

In general, higher temperatures can always be excluded when performing a systematic analysis of this type, due to their greater distance from the supposedly critical scaling regime, while lower temperatures should be included.
However, at lower temperatures, higher values of the chemical potential are expected at the chiral crossover, which may lie beyond the range of validity of the employed ansatz. 
Additionally, finite volume effects are expected to hinder LYZs from approaching the real axis for lower temperatures.
This may mostly affect the lowest temperatures, where the imaginary part of the LYZs are the smallest.
Therefore, we allow the fitting range to vary both at the lower and upper temperature boundaries, while always including the temperature range \SIrange{140}{150}{\MeV}.  

To account for all the systematic effects explored so far, $T_c$ is extrapolated using different combinations of the various methods, and the results are then combined into a cumulative distribution function (CDF).
This includes the three different approximated observables, all four ansatzes for the scaling relation, and all possible fits, as stated before.  
It should be noted that only a single type of Padé ([1,2]-Padé) was used, and therefore, systematic effects stemming from the order of the functional ansatz of the underlying observable are not accounted for.

\begin{figure}[ht]
\begin{minipage}{0.5\textwidth}
        \centering
    \includegraphics[height=5.60cm]{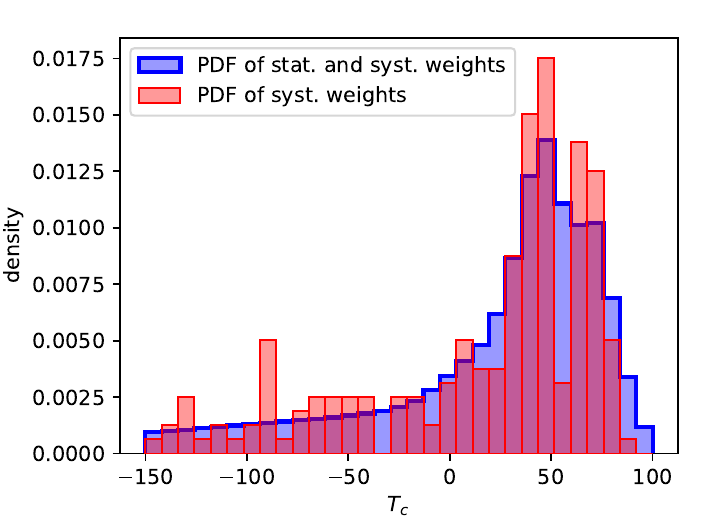}
\end{minipage}\begin{minipage}{0.5\textwidth}
    \centering
    \includegraphics[height=5.60cm]{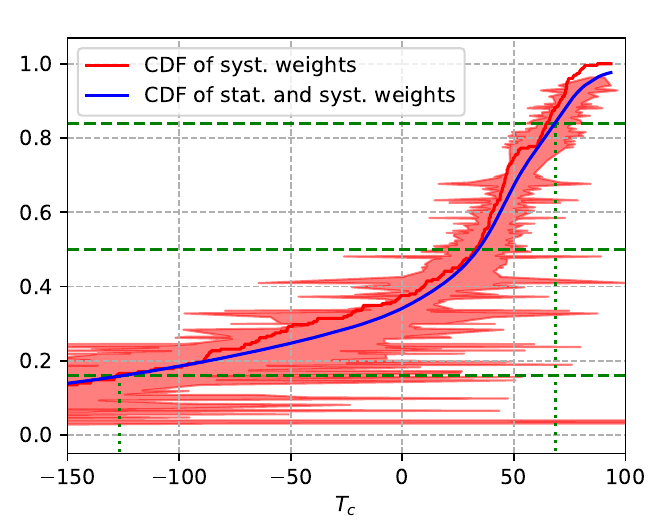}
\end{minipage}
    \caption{The probability distribution function (PDF) and the corresponding cumulative distribution function (CDF) derived from the results of the systematic analysis. The data selection is based on a p-value cut of 5\%, a fit range that included at least four temperatures, always incorporating the range \SIrange{140}{150}{\MeV}, as well as an artificial cut at \SI{-500}{\MeV}.
    The red curve represents the CDF based on systematic values alone, with the error band indicating statistical uncertainty. The blue curve shows the synthesized CDF combining both stat. and sys. uncertainties. The final result is $T_c^{50} = \SI{34}{\MeV}$, $T_c^{84} = \SI{67}{\MeV}$, and $T_c^{16} = \SI{-127}{\MeV}$.
}
    \label{fig:cdf}
\end{figure}

After applying a p-value cut of 5\%, the remaining $T_c$ values are included with flat weighting.
An additional arbitrary cut at $T_c = \SI{-500}{\MeV}$ was introduced to curb the extreme outliers, resulting in the removal of approximately 10\%. This cut is necessary, as the method used to determine $T_c$ calculating the intersection with the $x$-axis, yields a non-Gaussian distribution, which causes a skew in the CDF.
The resulting CDF and corresponding PDF is shown in \autoref{fig:cdf}. 
From it we have as median value of the critical temperature $T_c^{50} = \SI{34}{\MeV}$. The 84th percentile is located at $T_c^{84} = \SI{67}{\MeV}$ and the 16th percentile at $T_c^{16} = \SI{-127}{\MeV}$. The CDF reaches a value 34\% at a value of $T_c=0$.
\FloatBarrier
\section{Conclusions}
Padé approximations applied to the susceptibility, density, and pressure enabled the extraction of Lee-Yang zeros from a high-precision dataset. 
By employing different scaling observables and fit ranges, we extrapolated these results toward the Lee-Yang edge singularity and estimated the critical temperature.
Four ansatzes were used for the scaling observables, along with a carefully selected set of fit ranges, as detailed in the analysis section.  

The combination of all analyses, using a p-value cut of five percent and flat weighting, results in a CDF with a median of $T_c^{50} = \SI{34}{\MeV}$ with the 84th percentile located at $T_c^{84} = \SI{67}{\MeV}$. 
This analysis does not account for all systematic uncertainties but highlights the necessity of great care and consideration in making a CEP prediction.  
\vspace*{-0.0425cm}
\acknowledgments 
\vspace*{-0.0425cm}
\noindent This work is supported by the MKW NRW under the funding code NW21-024-A. 
Further funding was received from the DFG under the Project
No. 496127839. This work was also supported by the
Hungarian National Research, Development and Innovation
Office, NKFIH Grant No. KKP126769.
This work was also supported by the NKFIH excellence
grant TKP2021{\textunderscore}NKTA{\textunderscore}64.
This work is also supported by the Hungarian National Research,
Development and Innovation
Office under Project No. FK 147164.
The authors gratefully acknowledge the Gauss Centre for
Supercomputing e.V. (\url{www.gauss-centre.eu}) for funding
this project by providing computing time on the GCS
Supercomputer Juwels-Booster at Juelich Supercomputer
Centre.
We acknowledge the EuroHPC Joint Undertaking for awarding this project access to the EuroHPC supercomputer LUMI, hosted by CSC (Finland) and the LUMI consortium through a EuroHPC Extreme Access call.
D. P. is supported by the EKÖP-24 University Excellence Schorlarship Program of the Ministry for Culture and Innovation from the source of the National Research, Development and Innovation Fund.

\end{document}